\documentclass[%
 reprint,
superscriptaddress,
preprintnumbers,
 amsmath,amssymb,
 aps,
prl,
nobalancelastpage,
]{revtex4-2}

\usepackage{graphicx}
\usepackage{dcolumn}
\usepackage{bm}
\usepackage{hyperref}
\usepackage{comment}

\newcommand\al[1]{\begin{align}#1\end{align}}

\newcommand{\tr}{\operatorname{tr}}

\usepackage{xcolor}

\begin{document}

\preprint{RIKEN-iTHEMS-Report-26, STUPP-26-305}

\title{A gravitational realization of Courant-Hilbert deformations}

\author{Osamu Fukushima}
\email{ofukushi\_at\_mail.doshisha.ac.jp}
\affiliation{Faculty of Science and Engineering, Doshisha University, Kyoto 610-0394, Japan}
\affiliation{iTHEMS, RIKEN, Wako, Saitama 351-0198, Japan}

\author{Miyabi Kurihara}
\email{kurihara.m.038\_at\_ms.saitama-u.ac.jp}
\affiliation{Graduate School of Science and Engineering, Saitama University, 255 Shimo-Okubo, Sakura-ku, Saitama 338-8570, Japan}

\author{Takaki Matsumoto}
\email{takaki-matsumoto\_at\_ejs.seikei.ac.jp}
\affiliation{Seikei University, 3-3-1 Kichijoji-Kitamachi, Musashino-shi, Tokyo 180-8633, Japan}

\author{Kentaroh Yoshida}
\email{kenyoshida\_at\_mail.saitama-u.ac.jp}
\affiliation{Graduate School of Science and Engineering, Saitama University, 255 Shimo-Okubo, Sakura-ku, Saitama 338-8570, Japan}

\date{\today}

\begin{abstract}
Courant--Hilbert (CH) deformations unify broad classes of solvable stress-tensor flows including $T\bar{T}$ and root-$T\bar{T}$. However, a gravitational action that systematically yields such general deformations has remained unknown. In this Letter, we construct a two-dimensional massive gravity theory whose on-shell action reproduces the complete CH deformation. The gravity sector is defined via an arbitrary function of the eigenvalue ratio $y$ of the relative zweibein. To bypass the technical difficulty of directly eliminating the auxiliary zweibein, we reformulate the matter sector using a $2 \times 2$ spectral decomposition involving $y$ and a rank-one projector $K$. Sequentially solving the equations of motion---first for $K$ and subsequently for $y$---naturally maps the system onto the Russo--Townsend form of the CH flow. Our work extends Tolley's massive gravity formulation of $T\bar{T}$ flows to the CH landscape, providing a clear geometric mechanism for solvable stress-tensor flows.
\end{abstract}

\maketitle


\section{Introduction}

Understanding how physical systems deform under non-linear interactions is a central theme across quantum field theory, string theory, and statistical mechanics. In recent years, solvable deformations governed by operators composed of the energy-momentum tensor have emerged as a powerful paradigm for probing non-perturbative physics. At the forefront of this development is the $T\bar{T}$ deformation \cite{Smirnov:2016lqw,Cavaglia:2016oda}, which uniquely preserves the integrability of a two-dimensional theory while drastically altering its ultraviolet behavior, offering a promising bridge between field theory and quantum gravity. Indeed, turning on a $T\bar{T}$-like operator is dynamically realized by coupling the undeformed field theory to two-dimensional massive gravity \cite{Tolley:2019nmm,Babaei-Aghbolagh:2024hti,Tsolakidis:2024wut,Nix:2025plr}. 

Despite these successes, standard $T\bar{T}$ flows represent only a narrow slice of a much broader landscape of solvable deformations. A major step forward was recently achieved through the Courant-Hilbert (CH) deformation framework \cite{Babaei-Aghbolagh:2025uoz,Babaei-Aghbolagh:2025hlm,Fukushima:2025tlj,Fukushima:2026gan}, which generalizes stress-tensor flows into a unified picture. By reformulating stress-tensor deformations via non-linear partial differential equations \cite{Courant}, the CH framework accommodates a vast family of solvable flows, including both $T\bar{T}$ \cite{Smirnov:2016lqw,Cavaglia:2016oda} and root-$T\bar{T}$ \cite{Rodriguez:2021tcz,Babaei-Aghbolagh:2022uij,Ferko:2022cix,Babaei-Aghbolagh:2022leo,Tempo:2022ndz} deformations~\footnote{See also \cite{Babaei-Aghbolagh:2025cni,Babaei-Aghbolagh:2026vkm} for related studies in nonlinear electrodynamics.}. 
A complementary approach to integrable stress-tensor deformations
employs auxiliary fields. Auxiliary field sigma models (AFSMs)
generate an infinite family of integrable deformations of the
principal chiral model, including $T\bar T$ and root-$T\bar T$
\cite{Ferko:2024ali}. A scalar $\mu$-frame formulation relating
the original $\nu$-frame of AFSMs to the CH
formulations was subsequently developed in \cite{Baglioni:2025tsc}.
For related developments, see
\cite{Fukushima:2024nxm,Bielli:2024khq,Bielli:2024ach,Bielli:2024fnp,Bielli:2024oif,
Bielli:2025uiv,Ferko:2025bhv,Bielli:2026ggs,Bielli:2026vit}.

However, a fundamental piece of this mechanism remains missing: what gravitational theory corresponds to a general CH deformation? Uncovering this gravitational counterpart is essential to understanding how stress-tensor flows deform spacetime geometries beyond the standard $T\bar{T}$ regime.

In this Letter, we resolve this open problem by constructing a generalized two-dimensional massive gravity theory that yields generic CH flows upon setting the  gravitational fields on-shell. The gravitational action is defined via an arbitrary function of the eigenvalue ratio $y$ of the relative zweibein. To bypass the technical bottleneck of directly eliminating the auxiliary zweibein, we reformulate the matter sector using a $2\times 2$ spectral decomposition involving $y$ and a rank-one projector $K$ onto the timelike eigenspace. By sequentially solving the on-shell conditions—first for $K$ and then for $y$—we explicitly map the on-shell gravitational action onto the alternative CH construction introduced by Russo and Townsend \cite{Russo:2024ptw,Russo:2025fuc}. 
Our work extends Tolley's massive gravity proposal \cite{Tolley:2019nmm,Babaei-Aghbolagh:2024hti,Tsolakidis:2024wut,Nix:2025plr} to the CH landscape, providing a clear geometric mechanism for solvable stress-tensor deformations.

\medskip 

\noindent 
{\bf Note added:} After submitting our paper, a related paper \cite{Baglioni:2026mii} appeared. It explores integrable auxiliary field sigma models and their connection to metric approaches for $T\bar{T}$-like deformations. It is complementary to our work. 
It is interesting to investigate how their auxiliary field framework and master actions relate to the spectral decomposition presented in our approach.

\section{Geometric Realization}
\label{sec:gravitational-action}

Let us begin with the following action of a two-dimensional massive gravity, 
\begin{align}
  I[\phi,e,f]
  =
  I_{\mathrm{matter}}[\phi,e]
  +
  I_{\mathrm{grav}}[e,f]\,.
  \label{total_action}
\end{align}
The first part is the matter action in which the matter $\phi$ is coupled to the gravity described by the zweibein $e^a_\mu$\, where $\mu$ is the spacetime index and $a$ is the tangential index. The second part is characteristic to the massive gravity in which $e$ is coupled to another zweibein $f^a_\mu$\,. That is, the action (\ref{total_action}) contains the following two metrics, 
\begin{align}
  g_{\mu\nu} = \eta_{ab}e^a_\mu e^b_\nu,\qquad
  h_{\mu\nu} = \eta_{ab}f^a_\mu f^b_\nu.
  \label{metrics}
\end{align}
Throughout, $g^{\mu\nu}$ and $h^{\mu\nu}$ represent the inverses of $g_{\mu\nu}$ and $h_{\mu\nu}$, respectively. Unless noted otherwise, tensor indices are raised and lowered with the background metric $h$, while contractions with the auxiliary metric $g$ are made explicit. 

Our goal is to obtain the deformed matter action by eliminating the auxiliary zweibein $e$. We denote a solution of its equation of motion (eom) by $e^\ast(\phi,f)$:
\begin{align}
	\left.
	\frac{\delta I[\phi,e,f]}{\delta e^a_\mu}
	\right|_{e=e^\ast(\phi,f)} = 0.
	\label{auxiliary_eom_definition}
\end{align}
Substituting this solution into the total action, we obtain the effective action
\begin{align}
	I_{\mathrm{eff}}[\phi, f]
	:= I[\phi,e^\ast(\phi, f), f],
	\label{effective_action_definition}
\end{align}
which we identify with the deformed matter action:
\begin{align}
	I_{\mathrm{eff}}[\phi, f]
	= I_{\mathrm{deformed}}[\phi, f].
	\label{auxiliary_elimination_target}
\end{align}
To carry out this procedure, we first specify the gravitational action and the configuration space of $e$.

\section{Spectral decomposition}

We take the gravitational part of the action to be constructed algebraically from $e$ and $f$. It is convenient to introduce the relative matrix
\al
{
	A^\mu{}_\nu := (e^{-1})^\mu_a f^a_\nu\,.
}
The symmetric-zweibein condition should be imposed on $A$:
\begin{align}
    A^\mu{}_\rho A^\rho{}_\nu&=g^{\mu\rho}h_{\rho\nu}, \label{symmetric_vielbein}
\end{align}
which is equivalent to 
\al
{
	\label{symmetric_vielbein_equiv}
	h_{\mu\rho}A^\rho{}_\nu&=h_{\nu\rho}A^\rho{}_\mu.
} 
This condition means that $A$ is self-adjoint with respect to $h$: $h(Av,w)=h(v,Aw)$, where $h(v,w) := h_{\mu\nu}v^\mu w^\nu$. 

For simplicity, we suppose that $A$ has the two distinct real eigenvalues hereafter. 
To realize this supposition, we impose the following three conditions on $A$:
\begin{align}
  \det A&>0, \label{det_positive}\\
  (\tr A)^2-4\det A&>0, \label{real_eigenvalues}\\
  \tr A&>0. \label{trace_positive}
\end{align}
Condition \eqref{det_positive} fixes the relative orientation of $e$ and $f$.
Together, \eqref{det_positive} and \eqref{real_eigenvalues} imply that $A$ has two distinct real eigenvalues that are either both positive or both negative. 
Condition \eqref{trace_positive} selects the positive sign.

Due to the symmetric-zweibein condition \eqref{symmetric_vielbein}, $A$ has exactly three independent degrees of freedom. 
Thus, $A$ can be parametrized by two eigenvalues $\alpha_\text{t}$ and $\alpha_\text{s}$, whose corresponding eigenvectors are timelike and spacelike, respectively, and by the rank-one orthogonal projector $K$ onto its timelike eigenspace, as
\begin{align}
  A=\alpha_\text{t} K + \alpha_\text{s} (\mathbf{1}-K),
  \quad
  K^2=K,\quad
  \tr K=1.
  \label{spectral_decomposition}
\end{align}

\section{New gravitational action}

Invariance under diffeomorphisms and local Lorentz transformations acting simultaneously on $e$ and $f$, allows us to write \cite{Babaei-Aghbolagh:2024hti}
\al
{
	\label{general_grav_action}
	I_{\mathrm{grav}}[e,f] &= \int d^2x\,\det e\,B(A) \notag \\ 
    &= \int d^2x\,\det f\,\tilde{B}(A), 
}
where $B(A)$ is a scalar potential constructed from $A$, and we have defined $\tilde{B}(A) := B(A)/\det A$. For later convenience, we will work with $\tilde{B}(A)$ in the following analysis. 

It is helpful to parametrize the eigenvalues as
\al
{
	\label{parametrization}
	\alpha_{\mathrm{t}}=\alpha y,\qquad
	\alpha_{\mathrm{s}}=\alpha,\qquad
	y:=\frac{\alpha_{\mathrm{t}}}{\alpha_{\mathrm{s}}},
}
where $\alpha > 0$. Since the eigenvalues are positive and distinct, the allowed range of $y$ consists of the two intervals $0<y<1$ and $y>1$. The ratio $y$ can also be expressed directly in terms of $A$ as 
\begin{align}
  y=
  \frac{
    \tr A+\varepsilon\sqrt{(\tr A)^2-4\det A}
  }{
    \tr A-\varepsilon\sqrt{(\tr A)^2-4\det A}
  },
  \label{y_def_section_one}
\end{align}
where $\varepsilon:=\mathrm{sgn}(y-1)$, so that $\varepsilon=-1$ for $0<y<1$ and $\varepsilon=1$ for $y>1$.

Since $\tilde{B}(A)$ is a scalar function of $A$, $\tilde{B}(A)$ is written as 
\begin{align}
	\tilde{B}(A) = \tilde{B}(\alpha, y).
\end{align}
Note that under a Weyl transformation of $e$, the scale $\alpha$ changes while the ratio $y$ remains invariant. For simplicity, we suppose that the matter action is classically invariant under Weyl transformations of $e$, that is, it is independent of $\alpha$. Then, the eom for $\alpha$ therefore comes entirely from the gravitational action and is given by
\al
{
  \label{alpha_eom}
  \frac{\partial \tilde{B}(\alpha,y)}{\partial\alpha}=0.
}
We assume that this equation can be solved locally for $\alpha$, giving $\alpha=\alpha^\ast(y)$. After eliminating $\alpha$ using this solution, we obtain the reduced gravitational action
\al
{
	I_{\mathrm{grav}}[y, f]
	&:= \left.I_{\mathrm{grav}}[e,f]\right|_{\alpha=\alpha^\ast(y)} \notag \\ 
	&= -\int d^2x\,\det f\,\Omega(y),
    \label{I_grav-potential}
}
where we have introduced the following quantity: 
\al
{
	\Omega(y):= - \tilde{B}(\alpha^\ast(y),y).
}
If $\tilde{B}$ is independent of $\alpha$ from the outset, \eqref{alpha_eom} is identically satisfied, and we directly obtain 
$\Omega(y):= - \tilde{B}(y)$.

In the next section, we specify the matter action and complete the elimination of $e$. Then, the resulting action reproduces the alternative CH construction \cite{Russo:2024ptw,Russo:2025fuc}.

\section{Deriving CH deformations}
\label{sec:RT-form}

Let us start with the following sigma model action for the matter action, for simplicity, 
\begin{equation}
	I_{\rm matter}[\phi, e]
	= - \frac{1}{2} \int d^2x\,
	\det e\, g^{\mu\nu} G_{IJ}\partial_\mu\phi^I \partial_\nu \phi^J,
    \label{matter}
\end{equation}
where $G_{IJ}$ is the metric of the target spacetime and the indices $I$ and $J$ run from $1$ to $N$. We assume that $G_{IJ}$ is positive-definite. This action is obviously Weyl invariant at classical level.

It is helpful to define the symmetric tensor
\begin{equation}
	X_{\mu\nu} := G_{IJ}\partial_\mu\phi^I \partial_\nu \phi^J,
\end{equation}
and the associated mixed tensor $X^\mu{}_\nu = h^{\mu\rho} X_{\rho\nu}$. We use $X$ to denote the $2\times2$ matrix with indices $X^\mu{}_\nu$ hereafter. Since $X_{\mu\nu} = X_{\nu\mu}$, it follows that $h_{\mu\rho} X^\rho{}_\nu = h_{\nu\rho} X^\rho{}_\mu$. Thus, $X^\mu{}_\nu$ is self-adjoint with respect to $h_{\mu\nu}$.

In what follows, we assume that $X$ has two distinct real eigenvalues, namely,
\al
{
	\label{matter-condition}
	(\tr X)^2 - 4 \det X > 0.
}
Since $X$ is self-adjoint with respect to $h$, eigenvectors corresponding to its two distinct eigenvalues are orthogonal with respect to $h$. In two-dimensional Lorentzian signature, one of them is timelike and the other is spacelike. Moreover, the positive semidefiniteness of $X_{\mu\nu}$ implies that the eigenvalues associated with the timelike and spacelike eigenvectors are nonpositive and nonnegative, respectively. We denote them by $-2V$ and $2U$, respectively, where $U,V\geq0$. It then follows that 
\begin{align}
  \tr X=2(U-V),
  \qquad
  \det X=-4UV. 
  \label{X_trace_det}
\end{align}
In terms of $U$ and $V$, the condition \eqref{matter-condition} is equivalent to $U+V>0$.

By using \eqref{spectral_decomposition}, the following relations are derived: 

\begin{align}
  \frac{A^2}{\det A}
  =
  \frac{\alpha_\text{t}}{\alpha_\text{s}} K
  + \frac{\alpha_\text{s}}{\alpha_\text{t}} (\mathbf{1}-K)
  =
  y K + \frac{1}{y} (\mathbf{1}-K).
  \label{A2_detA}
\end{align}
As a result, the matter action is rewritten as 
\begin{align}
  I_{\rm matter}[\phi,y,K,f] 
  &=
  -\frac12 \int d^2x\,
	\det f \nonumber \\
  &\times\left[
    \frac{1}{y}\,\tr X
    +
    \left(y - \frac{1}{y}\right)\tr(KX)
  \right].
  \label{pcm_y_K}
\end{align}

The next task is to derive the eom for $K$. For fixed $y \neq 1$, the only $K$-dependent part is the second term in \eqref{pcm_y_K}. 
Since $y\neq 1$, the stationary condition for $K$ is $\tr(\delta K X)=0$ for every allowed $\delta K$. After some computation, one can see that this is equivalent to
\begin{align}
  [K,X] = 0\,.
  \label{K_variation_condition}
\end{align}

Let $K^\ast$ denote a rank-one projector onto a timelike direction that satisfies the eom for $K$. For any nonzero vector $v$ satisfying $K^\ast v = v$, the commutation relation $[K^\ast, X] = 0$ gives $K^\ast(Xv) = XK^\ast v = Xv$. Thus, $Xv$ belongs to the same one-dimensional subspace as $v$ and must be proportional to $v$. Therefore, $v$ is a timelike eigenvector of $X$. Since $X$ has a unique timelike eigenspace, $K^\ast$ is the spectral projector associated with the eigenvalue $-2V$, namely,
\al
{
	K^\ast = \frac{2U - X}{2U + 2V}.
}
Substituting this into \eqref{pcm_y_K}, we obtain
\begin{align}
  I_{\mathrm{matter}}[\phi,y,K^\ast,f]
  &=
  \int d^2x\,\det f
  \left(
    yV-\frac{U}{y}
  \right).
  \label{pcm_plus}
\end{align}

The final task is to evaluate the on-shell $y$-contribution. For this purpose, let us now add the gravity action \eqref{I_grav-potential},
\begin{align}
\! I[\phi,y,K^\ast,f]
  =
  \int d^2x\,\det f
  \left(
    yV-\frac{U}{y}-\Omega(y)
  \right).
  \label{action_plus}
\end{align}
Then, the eom for $y$ is
\begin{align}
  \Omega'(y)=V+\frac{U}{y^2}.
  \label{rt_constraint}
\end{align}
This is nothing but the alternative CH construction \cite{Russo:2024ptw,Russo:2025fuc}. At this stage, both disconnected regions $0 < y < 1$ and $ y > 1$ remain allowed.

Note that the right-hand side of \eqref{rt_constraint} is strictly positive for every $y > 0$, since $U, V \geq 0$ and $U+V>0$. Consequently, in either of the two allowed region $0 < y < 1$ and $y > 1$, the eom for $y$ can admit a solution only at a point where
\al
{
	\Omega'(y) > 0.
}

It is worth seeing the relation between the original CH and the alternative CH \cite{Russo:2024ptw,Russo:2025fuc}. Let us start from the original CH form, 
\begin{align}
  L
  =
  \ell(\tau)-\frac{2U}{\dot\ell(\tau)}, 
  \qquad 
  \tau=V+\frac{U}{\dot\ell(\tau)^2}. 
  \label{ch_form}
\end{align}
Then, performing the Legendre transformation 
\begin{equation}
  \Omega (y) = y \tau (y) - \ell (\tau (y)), \quad y = \dot\ell(\tau), \label{legendre}
\end{equation}
we obtain the Lagrangian 
\begin{equation}
L = y V   - \frac{U}{y} - \Omega(y).   
\end{equation}
Thus, we have reproduced the alternative CH form from the original CH one.

\section{Undeformed limit}
\label{sec:undeformed_limit}

We now address the undeformed limit ($\lambda \to 0$), where $\lambda$ represents the deformation parameter in $\Omega(y)$. At the off-shell level, $y$ and the matter sector are explicitly independent of $\lambda$. However, solving the eom for $y$ introduces a $\lambda$-dependence into both $y$ and the matter action. Consequently, the undeformed limit can be taken in two distinct ways: (i) on-shell (after integrating out $y$) and (ii) off-shell (before integrating out $y$).

In the on-shell scheme, the deformed matter action reduces smoothly to the undeformed one. This limit is conveniently analyzed in the original CH form \eqref{ch_form}, where the undeformed theory corresponds to
\begin{equation}
\ell_0(\tau) = \tau \quad \implies \quad 
y = \dot{\ell}_0 = 1, \quad \ddot{\ell}_0 = 0.
\label{undeformed_ch}
\end{equation}
The constraint in \eqref{ch_form} yields $\tau = V+U$, which reduces the Lagrangian to $L = V-U = -\tr X/2$, recovering the standard matter theory \eqref{matter}.
In this limit, the relation \eqref{legendre} means that the solution $y^*(\lambda) \to 1$ as $\lambda \to 0$. 
Thus, the undeformed configuration $y=1$ lies on the boundary of the domain where $y \neq 1$. When $y=1$, the eigenvalues degenerate ($\alpha_{\rm t} = \alpha_{\rm s}$), implying via \eqref{spectral_decomposition} that $A$ is proportional to the identity, i.e., $e=f$ up to Weyl transformation. 

For example, in the $T\bar{T}$-deformed theory~\cite{Babaei-Aghbolagh:2025hlm,Fukushima:2025tlj}, the CH function reads
\begin{equation}
\ell_{T\bar{T},\lambda}(\tau) 
= -\frac{1}{2\lambda}\sqrt{1-4\lambda\tau} 
+ \frac{1}{2\lambda}.
\label{ttbar_ch}
\end{equation}
Solving the eom \eqref{rt_constraint} yields the explicit on-shell value
\begin{equation}
y^* = \sqrt{\frac{1+4\lambda U}{1-4\lambda V}},
\label{TT-y}
\end{equation}
which generates the deformed gravitational potential $\Omega_{T\bar{T},\lambda}(y)$ and the effective Lagrangian:
\begin{align}
\Omega_{T\bar{T},\lambda}(y)=&\,
\frac{(1-y)^2}{4\lambda y},
\label{TTbar-potential}
\\
L_{T\bar{T},\lambda}(U,V,y^*)
=&\, \frac{1-\sqrt{(1+4\lambda U)(1-4\lambda V)}}{2\lambda}.
\end{align}
Taking the undeformed limit $\lambda \to 0$ in $L_{T\bar{T},\lambda}$ is straightforward and regular.

Conversely, taking the undeformed limit at the off-shell level is more subtle. Then, $y$ is treated as an independent variable from $\lambda$,
and thus the potential \eqref{TTbar-potential} diverges as $\lambda\to 0$.
This is because the Legendre transformation \eqref{legendre} becomes ill-defined in this limit because $\ddot{\ell}_0 = 0$. The non-invertibility of the map $\tau \mapsto y$ when $\ddot{\ell}=0$ underlies the apparent singular behavior in the alternative CH construction.

\section{Relation to Tolley's work}
\label{sec:TTbar}

It is worth seeing how our formulation incorporates the massive gravity description of the $T\bar{T}$-deformation in \cite{Tolley:2019nmm,Babaei-Aghbolagh:2024hti,Tsolakidis:2024wut,Nix:2025plr}.
In \cite{Tolley:2019nmm}, the total action for the $T\bar{T}$-deformation is given by
\begin{align}
    I_{\rm matter}[e,\phi]
    +
    \tilde{I}_{ T\bar{T}}[e,f]\,,
\end{align}
where the second term is defined as 
\begin{align}
    \tilde{I}_{ T\bar{T}}[e,f]
    :=&\,
    \int d^2x\,\frac{1}{2\lambda}\epsilon^{\mu\nu}\epsilon_{ab}(e^a_\mu-f^a_\mu)(e^b_\nu-f^b_\nu)
    \nonumber\\
    =&\,
    \int d^2x\,\det f \frac{1}{\lambda}\frac{(\alpha_{\rm s}-1)(\alpha_{\rm t}-1)}{\alpha_{\rm s}\alpha_{\rm t}}\,.
\end{align}
Here $\epsilon^{\mu\nu}$ and $\epsilon_{ab}$ are anti-symmetric tensors. 
Parametrizing the eigenvalues $(\alpha_{\rm s},\alpha_{\rm t})=(\alpha,\alpha y)$\,,
we obtain 
\begin{align}
    \tilde{I}_{T\bar{T}}[e,f]=&\, \int d^2x\, \det f 
    \frac{(\alpha-1)(\alpha y -1)}{\lambda \alpha^2y}\,.
\end{align}
Since $\tilde{I}_{T\bar{T}}$ is not Weyl invariant with respect to $e^a_\mu$, it has the explicit $\alpha$-dependence in contrast to the potential \eqref{TTbar-potential}.
In this case, we thus have to consider the eom
\begin{align}
    0=&\,
    \frac{\partial}{\partial \alpha}\frac{(\alpha-1)(\alpha y -1)}{\lambda \alpha^2y}
    =\frac{\alpha(1+y)-2}{\lambda\alpha^3 y},
\end{align}
and the solution is obtained as $\alpha^*=2/(1+y)$\,.
By substituting the solution, the Lagrangian reduces to
\begin{align}
    \frac{(\alpha^*-1)(\alpha^* y -1)}{\lambda \alpha^{*2}y}
    =-\frac{(1-y)^2}{4\lambda y}= -\Omega_{T\bar{T},\lambda}\,.
\end{align}
Thus, the action $\tilde{I}_{T\bar{T}}$ can be incorporated into the class of actions given by \eqref{general_grav_action}.

\section{Conclusion and Discussion}
\label{sec:conclusion}

In this Letter, we have constructed a generalized two-dimensional massive gravity theory that provides a gravitational realization of the CH deformation framework. By introducing a gravity action governed by an arbitrary function $\Omega(y)$ of the relative matrix eigenvalue ratio $y$, we bypassed the technical bottleneck of directly eliminating the zweibein $e_\mu^a$ through a $2\times 2$ matrix spectral decomposition of the relative matrix. Sequentially solving the eom for the rank-one timelike projector $K$ and subsequently for $y$ systematically reduces the system to the alternative CH construction. The detailed analysis including the causality will be announced in the near future \cite{FKMY}.

Our results extend Tolley's massive gravity formulation of $T\bar{T}$ flows \cite{Tolley:2019nmm,Babaei-Aghbolagh:2024hti,Tsolakidis:2024wut,Nix:2025plr} to the CH landscape, providing a concrete geometric mechanism for a vast class of solvable stress-tensor deformations. This extension opens up several intriguing directions for future research. First, while our construction focuses on two-dimensional spacetime, the core technique—utilizing a $2\times 2$ spectral decomposition based on the ratio of two independent eigenvalues—holds strong potential for extension to four-dimensional non-linear electrodynamics (4D NLED), where antisymmetric field strength tensors effectively reduce to two independent invariants. Developing a gravitational or auxiliary-field description for 4D NLED via this spectral approach would be a valuable endeavor. Second, investigating the quantum-mechanical properties of fluctuating geometries within this generalized massive gravity could shed light on the UV behavior and exact partition functions of CH-deformed quantum field theories (QFTs). Finally, generalizing this framework to non-Weyl-invariant matter actions or exploring higher-dimensional holographic realizations may reveal deeper connections between solvable deformations and spacetime geometry.

By revealing the gravitational mechanism behind the CH flow, this formulation provides not only a geometric foundation for solvable QFT deformations, but also a versatile algebraic framework that promises broad applications across classical and quantum spacetime physics.

\medskip 

\begin{acknowledgments}
We would like to thank Yuta Hasegawa for useful discussions at an earlier stage. 
The work of O.\,F.\ was supported by RIKEN Special Postdoctoral Researchers Program.
The work of K.~Y.~was supported in part by JSPS KAKENHI Grant No.~JP22H05115, 25K07313 and the Asahipen Hikari Foundation.
\end{acknowledgments}

\bibliographystyle{apsrev4-2}
\bibliography{auxiliary}

\begin{thebibliography}{34}%
\makeatletter
\providecommand \@ifxundefined [1]{%
 \@ifx{#1\undefined}
}%
\providecommand \@ifnum [1]{%
 \ifnum #1\expandafter \@firstoftwo
 \else \expandafter \@secondoftwo
 \fi
}%
\providecommand \@ifx [1]{%
 \ifx #1\expandafter \@firstoftwo
 \else \expandafter \@secondoftwo
 \fi
}%
\providecommand \natexlab [1]{#1}%
\providecommand \enquote  [1]{``#1''}%
\providecommand \bibnamefont  [1]{#1}%
\providecommand \bibfnamefont [1]{#1}%
\providecommand \citenamefont [1]{#1}%
\providecommand \href@noop [0]{\@secondoftwo}%
\providecommand \href [0]{\begingroup \@sanitize@url \@href}%
\providecommand \@href[1]{\@@startlink{#1}\@@href}%
\providecommand \@@href[1]{\endgroup#1\@@endlink}%
\providecommand \@sanitize@url [0]{\catcode `\\12\catcode `\$12\catcode `\&12\catcode `\#12\catcode `\^12\catcode `\_12\catcode `\%12\relax}%
\providecommand \@@startlink[1]{}%
\providecommand \@@endlink[0]{}%
\providecommand \url  [0]{\begingroup\@sanitize@url \@url }%
\providecommand \@url [1]{\endgroup\@href {#1}{\urlprefix }}%
\providecommand \urlprefix  [0]{URL }%
\providecommand \Eprint [0]{\href }%
\providecommand \doibase [0]{https://doi.org/}%
\providecommand \selectlanguage [0]{\@gobble}%
\providecommand \bibinfo  [0]{\@secondoftwo}%
\providecommand \bibfield  [0]{\@secondoftwo}%
\providecommand \translation [1]{[#1]}%
\providecommand \BibitemOpen [0]{}%
\providecommand \bibitemStop [0]{}%
\providecommand \bibitemNoStop [0]{.\EOS\space}%
\providecommand \EOS [0]{\spacefactor3000\relax}%
\providecommand \BibitemShut  [1]{\csname bibitem#1\endcsname}%
\let\auto@bib@innerbib\@empty
\bibitem [{\citenamefont {Smirnov}\ and\ \citenamefont {Zamolodchikov}(2017)}]{Smirnov:2016lqw}%
  \BibitemOpen
  \bibfield  {author} {\bibinfo {author} {\bibfnamefont {F.~A.}\ \bibnamefont {Smirnov}}\ and\ \bibinfo {author} {\bibfnamefont {A.~B.}\ \bibnamefont {Zamolodchikov}},\ }\href {https://doi.org/10.1016/j.nuclphysb.2016.12.014} {\bibfield  {journal} {\bibinfo  {journal} {Nucl. Phys. B}\ }\textbf {\bibinfo {volume} {915}},\ \bibinfo {pages} {363} (\bibinfo {year} {2017})},\ \Eprint {https://arxiv.org/abs/1608.05499} {arXiv:1608.05499 [hep-th]} \BibitemShut {NoStop}%
\bibitem [{\citenamefont {Cavagli\`a}\ \emph {et~al.}(2016)\citenamefont {Cavagli\`a}, \citenamefont {Negro}, \citenamefont {Sz\'ecs\'enyi},\ and\ \citenamefont {Tateo}}]{Cavaglia:2016oda}%
  \BibitemOpen
  \bibfield  {author} {\bibinfo {author} {\bibfnamefont {A.}~\bibnamefont {Cavagli\`a}}, \bibinfo {author} {\bibfnamefont {S.}~\bibnamefont {Negro}}, \bibinfo {author} {\bibfnamefont {I.~M.}\ \bibnamefont {Sz\'ecs\'enyi}},\ and\ \bibinfo {author} {\bibfnamefont {R.}~\bibnamefont {Tateo}},\ }\href {https://doi.org/10.1007/JHEP10(2016)112} {\bibfield  {journal} {\bibinfo  {journal} {JHEP}\ }\textbf {\bibinfo {volume} {10}},\ \bibinfo {pages} {112}},\ \Eprint {https://arxiv.org/abs/1608.05534} {arXiv:1608.05534 [hep-th]} \BibitemShut {NoStop}%
\bibitem [{\citenamefont {Tolley}(2020)}]{Tolley:2019nmm}%
  \BibitemOpen
  \bibfield  {author} {\bibinfo {author} {\bibfnamefont {A.~J.}\ \bibnamefont {Tolley}},\ }\href {https://doi.org/10.1007/JHEP06(2020)050} {\bibfield  {journal} {\bibinfo  {journal} {JHEP}\ }\textbf {\bibinfo {volume} {06}},\ \bibinfo {pages} {050}},\ \Eprint {https://arxiv.org/abs/1911.06142} {arXiv:1911.06142 [hep-th]} \BibitemShut {NoStop}%
\bibitem [{\citenamefont {Babaei-Aghbolagh}\ \emph {et~al.}(2024)\citenamefont {Babaei-Aghbolagh}, \citenamefont {He}, \citenamefont {Morone}, \citenamefont {Ouyang},\ and\ \citenamefont {Tateo}}]{Babaei-Aghbolagh:2024hti}%
  \BibitemOpen
  \bibfield  {author} {\bibinfo {author} {\bibfnamefont {H.}~\bibnamefont {Babaei-Aghbolagh}}, \bibinfo {author} {\bibfnamefont {S.}~\bibnamefont {He}}, \bibinfo {author} {\bibfnamefont {T.}~\bibnamefont {Morone}}, \bibinfo {author} {\bibfnamefont {H.}~\bibnamefont {Ouyang}},\ and\ \bibinfo {author} {\bibfnamefont {R.}~\bibnamefont {Tateo}},\ }\href {https://doi.org/10.1103/PhysRevLett.133.111602} {\bibfield  {journal} {\bibinfo  {journal} {Phys. Rev. Lett.}\ }\textbf {\bibinfo {volume} {133}},\ \bibinfo {pages} {111602} (\bibinfo {year} {2024})},\ \Eprint {https://arxiv.org/abs/2405.03465} {arXiv:2405.03465 [hep-th]} \BibitemShut {NoStop}%
\bibitem [{\citenamefont {Tsolakidis}(2024)}]{Tsolakidis:2024wut}%
  \BibitemOpen
  \bibfield  {author} {\bibinfo {author} {\bibfnamefont {E.}~\bibnamefont {Tsolakidis}},\ }\href {https://doi.org/10.1007/JHEP09(2024)167} {\bibfield  {journal} {\bibinfo  {journal} {JHEP}\ }\textbf {\bibinfo {volume} {09}},\ \bibinfo {pages} {167}},\ \Eprint {https://arxiv.org/abs/2405.07967} {arXiv:2405.07967 [hep-th]} \BibitemShut {NoStop}%
\bibitem [{\citenamefont {Nix}\ and\ \citenamefont {Tsolakidis}(2026)}]{Nix:2025plr}%
  \BibitemOpen
  \bibfield  {author} {\bibinfo {author} {\bibfnamefont {A.}~\bibnamefont {Nix}}\ and\ \bibinfo {author} {\bibfnamefont {E.}~\bibnamefont {Tsolakidis}},\ }\href {https://doi.org/10.1007/JHEP09(2026)199} {\bibfield  {journal} {\bibinfo  {journal} {JHEP}\ }\textbf {\bibinfo {volume} {09}},\ \bibinfo {pages} {199}},\ \Eprint {https://arxiv.org/abs/2512.23533} {arXiv:2512.23533 [hep-th]} \BibitemShut {NoStop}%
\bibitem [{\citenamefont {Babaei-Aghbolagh}\ \emph {et~al.}(2025{\natexlab{a}})\citenamefont {Babaei-Aghbolagh}, \citenamefont {Chen},\ and\ \citenamefont {He}}]{Babaei-Aghbolagh:2025uoz}%
  \BibitemOpen
  \bibfield  {author} {\bibinfo {author} {\bibfnamefont {H.}~\bibnamefont {Babaei-Aghbolagh}}, \bibinfo {author} {\bibfnamefont {B.}~\bibnamefont {Chen}},\ and\ \bibinfo {author} {\bibfnamefont {S.}~\bibnamefont {He}},\ }\href {https://doi.org/10.1103/1r4p-3r5q} {\bibfield  {journal} {\bibinfo  {journal} {Phys. Rev. D}\ }\textbf {\bibinfo {volume} {112}},\ \bibinfo {pages} {L101702} (\bibinfo {year} {2025}{\natexlab{a}})},\ \Eprint {https://arxiv.org/abs/2507.22808} {arXiv:2507.22808 [hep-th]} \BibitemShut {NoStop}%
\bibitem [{\citenamefont {Babaei-Aghbolagh}\ \emph {et~al.}(2026{\natexlab{a}})\citenamefont {Babaei-Aghbolagh}, \citenamefont {Chen},\ and\ \citenamefont {He}}]{Babaei-Aghbolagh:2025hlm}%
  \BibitemOpen
  \bibfield  {author} {\bibinfo {author} {\bibfnamefont {H.}~\bibnamefont {Babaei-Aghbolagh}}, \bibinfo {author} {\bibfnamefont {B.}~\bibnamefont {Chen}},\ and\ \bibinfo {author} {\bibfnamefont {S.}~\bibnamefont {He}},\ }\href {https://doi.org/10.1007/JHEP01(2026)108} {\bibfield  {journal} {\bibinfo  {journal} {JHEP}\ }\textbf {\bibinfo {volume} {01}},\ \bibinfo {pages} {108}},\ \Eprint {https://arxiv.org/abs/2509.17075} {arXiv:2509.17075 [hep-th]} \BibitemShut {NoStop}%
\bibitem [{\citenamefont {Fukushima}\ \emph {et~al.}(2026{\natexlab{a}})\citenamefont {Fukushima}, \citenamefont {Matsumoto},\ and\ \citenamefont {Yoshida}}]{Fukushima:2025tlj}%
  \BibitemOpen
  \bibfield  {author} {\bibinfo {author} {\bibfnamefont {O.}~\bibnamefont {Fukushima}}, \bibinfo {author} {\bibfnamefont {T.}~\bibnamefont {Matsumoto}},\ and\ \bibinfo {author} {\bibfnamefont {K.}~\bibnamefont {Yoshida}},\ }\href {https://doi.org/10.1007/JHEP01(2026)122} {\bibfield  {journal} {\bibinfo  {journal} {JHEP}\ }\textbf {\bibinfo {volume} {01}},\ \bibinfo {pages} {122}},\ \Eprint {https://arxiv.org/abs/2509.22080} {arXiv:2509.22080 [hep-th]} \BibitemShut {NoStop}%
\bibitem [{\citenamefont {Fukushima}\ \emph {et~al.}(2026{\natexlab{b}})\citenamefont {Fukushima}, \citenamefont {Matsumoto},\ and\ \citenamefont {Yoshida}}]{Fukushima:2026gan}%
  \BibitemOpen
  \bibfield  {author} {\bibinfo {author} {\bibfnamefont {O.}~\bibnamefont {Fukushima}}, \bibinfo {author} {\bibfnamefont {T.}~\bibnamefont {Matsumoto}},\ and\ \bibinfo {author} {\bibfnamefont {K.}~\bibnamefont {Yoshida}},\ }\href {https://doi.org/10.1007/JHEP06(2026)084} {\bibfield  {journal} {\bibinfo  {journal} {JHEP}\ }\textbf {\bibinfo {volume} {06}},\ \bibinfo {pages} {084}},\ \Eprint {https://arxiv.org/abs/2602.04662} {arXiv:2602.04662 [hep-th]} \BibitemShut {NoStop}%
\bibitem [{\citenamefont {Courant}\ and\ \citenamefont {Hilbert}(1962)}]{Courant}%
  \BibitemOpen
  \bibfield  {author} {\bibinfo {author} {\bibfnamefont {R.}~\bibnamefont {Courant}}\ and\ \bibinfo {author} {\bibfnamefont {D.}~\bibnamefont {Hilbert}},\ }\href@noop {} {\emph {\bibinfo {title} {Methods of Mathematical Physics}}},\ Vol.~\bibinfo {volume} {II}\ (\bibinfo  {publisher} {Wiley Interscience},\ \bibinfo {year} {1962})\BibitemShut {NoStop}%
\bibitem [{\citenamefont {Rodr{\'\i}guez}\ \emph {et~al.}(2021)\citenamefont {Rodr{\'\i}guez}, \citenamefont {Tempo},\ and\ \citenamefont {Troncoso}}]{Rodriguez:2021tcz}%
  \BibitemOpen
  \bibfield  {author} {\bibinfo {author} {\bibfnamefont {P.}~\bibnamefont {Rodr{\'\i}guez}}, \bibinfo {author} {\bibfnamefont {D.}~\bibnamefont {Tempo}},\ and\ \bibinfo {author} {\bibfnamefont {R.}~\bibnamefont {Troncoso}},\ }\href {https://doi.org/10.1007/JHEP11(2021)133} {\bibfield  {journal} {\bibinfo  {journal} {JHEP}\ }\textbf {\bibinfo {volume} {11}},\ \bibinfo {pages} {133}},\ \Eprint {https://arxiv.org/abs/2106.09750} {arXiv:2106.09750 [hep-th]} \BibitemShut {NoStop}%
\bibitem [{\citenamefont {Babaei-Aghbolagh}\ \emph {et~al.}(2022{\natexlab{a}})\citenamefont {Babaei-Aghbolagh}, \citenamefont {Velni}, \citenamefont {Yekta},\ and\ \citenamefont {Mohammadzadeh}}]{Babaei-Aghbolagh:2022uij}%
  \BibitemOpen
  \bibfield  {author} {\bibinfo {author} {\bibfnamefont {H.}~\bibnamefont {Babaei-Aghbolagh}}, \bibinfo {author} {\bibfnamefont {K.~B.}\ \bibnamefont {Velni}}, \bibinfo {author} {\bibfnamefont {D.~M.}\ \bibnamefont {Yekta}},\ and\ \bibinfo {author} {\bibfnamefont {H.}~\bibnamefont {Mohammadzadeh}},\ }\href {https://doi.org/10.1016/j.physletb.2022.137079} {\bibfield  {journal} {\bibinfo  {journal} {Phys. Lett. B}\ }\textbf {\bibinfo {volume} {829}},\ \bibinfo {pages} {137079} (\bibinfo {year} {2022}{\natexlab{a}})},\ \Eprint {https://arxiv.org/abs/2202.11156} {arXiv:2202.11156 [hep-th]} \BibitemShut {NoStop}%
\bibitem [{\citenamefont {Ferko}\ \emph {et~al.}(2022)\citenamefont {Ferko}, \citenamefont {Sfondrini}, \citenamefont {Smith},\ and\ \citenamefont {Tartaglino-Mazzucchelli}}]{Ferko:2022cix}%
  \BibitemOpen
  \bibfield  {author} {\bibinfo {author} {\bibfnamefont {C.}~\bibnamefont {Ferko}}, \bibinfo {author} {\bibfnamefont {A.}~\bibnamefont {Sfondrini}}, \bibinfo {author} {\bibfnamefont {L.}~\bibnamefont {Smith}},\ and\ \bibinfo {author} {\bibfnamefont {G.}~\bibnamefont {Tartaglino-Mazzucchelli}},\ }\href {https://doi.org/10.1103/PhysRevLett.129.201604} {\bibfield  {journal} {\bibinfo  {journal} {Phys. Rev. Lett.}\ }\textbf {\bibinfo {volume} {129}},\ \bibinfo {pages} {201604} (\bibinfo {year} {2022})},\ \Eprint {https://arxiv.org/abs/2206.10515} {arXiv:2206.10515 [hep-th]} \BibitemShut {NoStop}%
\bibitem [{\citenamefont {Babaei-Aghbolagh}\ \emph {et~al.}(2022{\natexlab{b}})\citenamefont {Babaei-Aghbolagh}, \citenamefont {Babaei~Velni}, \citenamefont {Mahdavian~Yekta},\ and\ \citenamefont {Mohammadzadeh}}]{Babaei-Aghbolagh:2022leo}%
  \BibitemOpen
  \bibfield  {author} {\bibinfo {author} {\bibfnamefont {H.}~\bibnamefont {Babaei-Aghbolagh}}, \bibinfo {author} {\bibfnamefont {K.}~\bibnamefont {Babaei~Velni}}, \bibinfo {author} {\bibfnamefont {D.}~\bibnamefont {Mahdavian~Yekta}},\ and\ \bibinfo {author} {\bibfnamefont {H.}~\bibnamefont {Mohammadzadeh}},\ }\href {https://doi.org/10.1103/PhysRevD.106.086022} {\bibfield  {journal} {\bibinfo  {journal} {Phys. Rev. D}\ }\textbf {\bibinfo {volume} {106}},\ \bibinfo {pages} {086022} (\bibinfo {year} {2022}{\natexlab{b}})},\ \Eprint {https://arxiv.org/abs/2206.12677} {arXiv:2206.12677 [hep-th]} \BibitemShut {NoStop}%
\bibitem [{\citenamefont {Tempo}\ and\ \citenamefont {Troncoso}(2022)}]{Tempo:2022ndz}%
  \BibitemOpen
  \bibfield  {author} {\bibinfo {author} {\bibfnamefont {D.}~\bibnamefont {Tempo}}\ and\ \bibinfo {author} {\bibfnamefont {R.}~\bibnamefont {Troncoso}},\ }\href {https://doi.org/10.1007/JHEP12(2022)129} {\bibfield  {journal} {\bibinfo  {journal} {JHEP}\ }\textbf {\bibinfo {volume} {12}},\ \bibinfo {pages} {129}},\ \Eprint {https://arxiv.org/abs/2210.00059} {arXiv:2210.00059 [hep-th]} \BibitemShut {NoStop}%
\bibitem [{Note1()}]{Note1}%
  \BibitemOpen
  \bibinfo {note} {See also \cite {Babaei-Aghbolagh:2025cni,Babaei-Aghbolagh:2026vkm} for related studies in nonlinear electrodynamics.}\BibitemShut {Stop}%
\bibitem [{\citenamefont {Ferko}\ and\ \citenamefont {Smith}(2024)}]{Ferko:2024ali}%
  \BibitemOpen
  \bibfield  {author} {\bibinfo {author} {\bibfnamefont {C.}~\bibnamefont {Ferko}}\ and\ \bibinfo {author} {\bibfnamefont {L.}~\bibnamefont {Smith}},\ }\href {https://doi.org/10.1103/PhysRevLett.133.131602} {\bibfield  {journal} {\bibinfo  {journal} {Phys. Rev. Lett.}\ }\textbf {\bibinfo {volume} {133}},\ \bibinfo {pages} {131602} (\bibinfo {year} {2024})},\ \Eprint {https://arxiv.org/abs/2405.05899} {arXiv:2405.05899 [hep-th]} \BibitemShut {NoStop}%
\bibitem [{\citenamefont {Baglioni}\ \emph {et~al.}(2026{\natexlab{a}})\citenamefont {Baglioni}, \citenamefont {Bielli}, \citenamefont {Galli},\ and\ \citenamefont {Tartaglino-Mazzucchelli}}]{Baglioni:2025tsc}%
  \BibitemOpen
  \bibfield  {author} {\bibinfo {author} {\bibfnamefont {N.}~\bibnamefont {Baglioni}}, \bibinfo {author} {\bibfnamefont {D.}~\bibnamefont {Bielli}}, \bibinfo {author} {\bibfnamefont {M.}~\bibnamefont {Galli}},\ and\ \bibinfo {author} {\bibfnamefont {G.}~\bibnamefont {Tartaglino-Mazzucchelli}},\ }\href {https://doi.org/10.1007/JHEP07(2026)039} {\bibfield  {journal} {\bibinfo  {journal} {JHEP}\ }\textbf {\bibinfo {volume} {07}},\ \bibinfo {pages} {039}},\ \Eprint {https://arxiv.org/abs/2512.21982} {arXiv:2512.21982 [hep-th]} \BibitemShut {NoStop}%
\bibitem [{\citenamefont {Fukushima}\ and\ \citenamefont {Yoshida}(2025)}]{Fukushima:2024nxm}%
  \BibitemOpen
  \bibfield  {author} {\bibinfo {author} {\bibfnamefont {O.}~\bibnamefont {Fukushima}}\ and\ \bibinfo {author} {\bibfnamefont {K.}~\bibnamefont {Yoshida}},\ }\href {https://doi.org/10.1007/JHEP09(2025)001} {\bibfield  {journal} {\bibinfo  {journal} {JHEP}\ }\textbf {\bibinfo {volume} {09}},\ \bibinfo {pages} {001}},\ \Eprint {https://arxiv.org/abs/2407.02204} {arXiv:2407.02204 [hep-th]} \BibitemShut {NoStop}%
\bibitem [{\citenamefont {Bielli}\ \emph {et~al.}(2025{\natexlab{a}})\citenamefont {Bielli}, \citenamefont {Ferko}, \citenamefont {Smith},\ and\ \citenamefont {Tartaglino-Mazzucchelli}}]{Bielli:2024khq}%
  \BibitemOpen
  \bibfield  {author} {\bibinfo {author} {\bibfnamefont {D.}~\bibnamefont {Bielli}}, \bibinfo {author} {\bibfnamefont {C.}~\bibnamefont {Ferko}}, \bibinfo {author} {\bibfnamefont {L.}~\bibnamefont {Smith}},\ and\ \bibinfo {author} {\bibfnamefont {G.}~\bibnamefont {Tartaglino-Mazzucchelli}},\ }\href {https://doi.org/10.1103/PhysRevLett.134.101601} {\bibfield  {journal} {\bibinfo  {journal} {Phys. Rev. Lett.}\ }\textbf {\bibinfo {volume} {134}},\ \bibinfo {pages} {101601} (\bibinfo {year} {2025}{\natexlab{a}})},\ \Eprint {https://arxiv.org/abs/2407.11636} {arXiv:2407.11636 [hep-th]} \BibitemShut {NoStop}%
\bibitem [{\citenamefont {Bielli}\ \emph {et~al.}(2025{\natexlab{b}})\citenamefont {Bielli}, \citenamefont {Ferko}, \citenamefont {Smith},\ and\ \citenamefont {Tartaglino-Mazzucchelli}}]{Bielli:2024ach}%
  \BibitemOpen
  \bibfield  {author} {\bibinfo {author} {\bibfnamefont {D.}~\bibnamefont {Bielli}}, \bibinfo {author} {\bibfnamefont {C.}~\bibnamefont {Ferko}}, \bibinfo {author} {\bibfnamefont {L.}~\bibnamefont {Smith}},\ and\ \bibinfo {author} {\bibfnamefont {G.}~\bibnamefont {Tartaglino-Mazzucchelli}},\ }\href {https://doi.org/10.1103/PhysRevD.111.066010} {\bibfield  {journal} {\bibinfo  {journal} {Phys. Rev. D}\ }\textbf {\bibinfo {volume} {111}},\ \bibinfo {pages} {066010} (\bibinfo {year} {2025}{\natexlab{b}})},\ \Eprint {https://arxiv.org/abs/2407.16338} {arXiv:2407.16338 [hep-th]} \BibitemShut {NoStop}%
\bibitem [{\citenamefont {Bielli}\ \emph {et~al.}(2025{\natexlab{c}})\citenamefont {Bielli}, \citenamefont {Ferko}, \citenamefont {Smith},\ and\ \citenamefont {Tartaglino-Mazzucchelli}}]{Bielli:2024fnp}%
  \BibitemOpen
  \bibfield  {author} {\bibinfo {author} {\bibfnamefont {D.}~\bibnamefont {Bielli}}, \bibinfo {author} {\bibfnamefont {C.}~\bibnamefont {Ferko}}, \bibinfo {author} {\bibfnamefont {L.}~\bibnamefont {Smith}},\ and\ \bibinfo {author} {\bibfnamefont {G.}~\bibnamefont {Tartaglino-Mazzucchelli}},\ }\href {https://doi.org/10.1007/JHEP05(2025)223} {\bibfield  {journal} {\bibinfo  {journal} {JHEP}\ }\textbf {\bibinfo {volume} {05}},\ \bibinfo {pages} {223}},\ \Eprint {https://arxiv.org/abs/2408.09714} {arXiv:2408.09714 [hep-th]} \BibitemShut {NoStop}%
\bibitem [{\citenamefont {Bielli}\ \emph {et~al.}(2025{\natexlab{d}})\citenamefont {Bielli}, \citenamefont {Ferko}, \citenamefont {Smith},\ and\ \citenamefont {Tartaglino-Mazzucchelli}}]{Bielli:2024oif}%
  \BibitemOpen
  \bibfield  {author} {\bibinfo {author} {\bibfnamefont {D.}~\bibnamefont {Bielli}}, \bibinfo {author} {\bibfnamefont {C.}~\bibnamefont {Ferko}}, \bibinfo {author} {\bibfnamefont {L.}~\bibnamefont {Smith}},\ and\ \bibinfo {author} {\bibfnamefont {G.}~\bibnamefont {Tartaglino-Mazzucchelli}},\ }\href {https://doi.org/10.1007/JHEP01(2025)096} {\bibfield  {journal} {\bibinfo  {journal} {JHEP}\ }\textbf {\bibinfo {volume} {01}},\ \bibinfo {pages} {096}},\ \Eprint {https://arxiv.org/abs/2409.05704} {arXiv:2409.05704 [hep-th]} \BibitemShut {NoStop}%
\bibitem [{\citenamefont {Bielli}\ \emph {et~al.}(2025{\natexlab{e}})\citenamefont {Bielli}, \citenamefont {Ferko}, \citenamefont {Galli},\ and\ \citenamefont {Tartaglino-Mazzucchelli}}]{Bielli:2025uiv}%
  \BibitemOpen
  \bibfield  {author} {\bibinfo {author} {\bibfnamefont {D.}~\bibnamefont {Bielli}}, \bibinfo {author} {\bibfnamefont {C.}~\bibnamefont {Ferko}}, \bibinfo {author} {\bibfnamefont {M.}~\bibnamefont {Galli}},\ and\ \bibinfo {author} {\bibfnamefont {G.}~\bibnamefont {Tartaglino-Mazzucchelli}},\ }\href {https://doi.org/10.1007/JHEP08(2025)078} {\bibfield  {journal} {\bibinfo  {journal} {JHEP}\ }\textbf {\bibinfo {volume} {08}},\ \bibinfo {pages} {078}},\ \Eprint {https://arxiv.org/abs/2504.17294} {arXiv:2504.17294 [hep-th]} \BibitemShut {NoStop}%
\bibitem [{\citenamefont {Ferko}\ \emph {et~al.}(2026)\citenamefont {Ferko}, \citenamefont {Galli}, \citenamefont {Huang},\ and\ \citenamefont {Tartaglino-Mazzucchelli}}]{Ferko:2025bhv}%
  \BibitemOpen
  \bibfield  {author} {\bibinfo {author} {\bibfnamefont {C.}~\bibnamefont {Ferko}}, \bibinfo {author} {\bibfnamefont {M.}~\bibnamefont {Galli}}, \bibinfo {author} {\bibfnamefont {Z.}~\bibnamefont {Huang}},\ and\ \bibinfo {author} {\bibfnamefont {G.}~\bibnamefont {Tartaglino-Mazzucchelli}},\ }\href {https://doi.org/10.1007/JHEP03(2026)144} {\bibfield  {journal} {\bibinfo  {journal} {JHEP}\ }\textbf {\bibinfo {volume} {03}},\ \bibinfo {pages} {144}},\ \Eprint {https://arxiv.org/abs/2509.05081} {arXiv:2509.05081 [hep-th]} \BibitemShut {NoStop}%
\bibitem [{\citenamefont {Bielli}\ \emph {et~al.}(2026)\citenamefont {Bielli}, \citenamefont {Ferko}, \citenamefont {Galli},\ and\ \citenamefont {Tartaglino-Mazzucchelli}}]{Bielli:2026ggs}%
  \BibitemOpen
  \bibfield  {author} {\bibinfo {author} {\bibfnamefont {D.}~\bibnamefont {Bielli}}, \bibinfo {author} {\bibfnamefont {C.}~\bibnamefont {Ferko}}, \bibinfo {author} {\bibfnamefont {M.}~\bibnamefont {Galli}},\ and\ \bibinfo {author} {\bibfnamefont {G.}~\bibnamefont {Tartaglino-Mazzucchelli}},\ }\href@noop {} {\  (\bibinfo {year} {2026})},\ \Eprint {https://arxiv.org/abs/2605.18213} {arXiv:2605.18213 [hep-th]} \BibitemShut {NoStop}%
\bibitem [{\citenamefont {Bielli}\ and\ \citenamefont {Ces{\`a}ro}(2026)}]{Bielli:2026vit}%
  \BibitemOpen
  \bibfield  {author} {\bibinfo {author} {\bibfnamefont {D.}~\bibnamefont {Bielli}}\ and\ \bibinfo {author} {\bibfnamefont {M.}~\bibnamefont {Ces{\`a}ro}},\ }\href@noop {} {\  (\bibinfo {year} {2026})},\ \Eprint {https://arxiv.org/abs/2606.30717} {arXiv:2606.30717 [hep-th]} \BibitemShut {NoStop}%
\bibitem [{\citenamefont {Russo}\ and\ \citenamefont {Townsend}(2024)}]{Russo:2024ptw}%
  \BibitemOpen
  \bibfield  {author} {\bibinfo {author} {\bibfnamefont {J.~G.}\ \bibnamefont {Russo}}\ and\ \bibinfo {author} {\bibfnamefont {P.~K.}\ \bibnamefont {Townsend}},\ }\href {https://doi.org/10.1007/JHEP09(2024)107} {\bibfield  {journal} {\bibinfo  {journal} {JHEP}\ }\textbf {\bibinfo {volume} {09}},\ \bibinfo {pages} {107}},\ \Eprint {https://arxiv.org/abs/2407.02577} {arXiv:2407.02577 [hep-th]} \BibitemShut {NoStop}%
\bibitem [{\citenamefont {Russo}\ and\ \citenamefont {Townsend}(2025)}]{Russo:2025fuc}%
  \BibitemOpen
  \bibfield  {author} {\bibinfo {author} {\bibfnamefont {J.~G.}\ \bibnamefont {Russo}}\ and\ \bibinfo {author} {\bibfnamefont {P.~K.}\ \bibnamefont {Townsend}},\ }\href {https://doi.org/10.1007/JHEP10(2025)120} {\bibfield  {journal} {\bibinfo  {journal} {JHEP}\ }\textbf {\bibinfo {volume} {10}},\ \bibinfo {pages} {120}},\ \Eprint {https://arxiv.org/abs/2505.08869} {arXiv:2505.08869 [hep-th]} \BibitemShut {NoStop}%
\bibitem [{\citenamefont {Baglioni}\ \emph {et~al.}(2026{\natexlab{b}})\citenamefont {Baglioni}, \citenamefont {Bielli}, \citenamefont {Shah},\ and\ \citenamefont {Tartaglino-Mazzucchelli}}]{Baglioni:2026mii}%
  \BibitemOpen
  \bibfield  {author} {\bibinfo {author} {\bibfnamefont {N.}~\bibnamefont {Baglioni}}, \bibinfo {author} {\bibfnamefont {D.}~\bibnamefont {Bielli}}, \bibinfo {author} {\bibfnamefont {P.}~\bibnamefont {Shah}},\ and\ \bibinfo {author} {\bibfnamefont {G.}~\bibnamefont {Tartaglino-Mazzucchelli}},\ }\href@noop {} {\  (\bibinfo {year} {2026}{\natexlab{b}})},\ \Eprint {https://arxiv.org/abs/2610.13501} {arXiv:2610.13501 [hep-th]} \BibitemShut {NoStop}%
\bibitem [{FKM()}]{FKMY}%
  \BibitemOpen
  \href@noop {} {}\bibinfo {note} {O.~Fukushima, M.~Kurihara, T.~Matsumoto and K.~Yoshida, in preparation.}\BibitemShut {Stop}%
\bibitem [{\citenamefont {Babaei-Aghbolagh}\ \emph {et~al.}(2025{\natexlab{b}})\citenamefont {Babaei-Aghbolagh}, \citenamefont {Babaei~Velni}, \citenamefont {He},\ and\ \citenamefont {Pezhman}}]{Babaei-Aghbolagh:2025cni}%
  \BibitemOpen
  \bibfield  {author} {\bibinfo {author} {\bibfnamefont {H.}~\bibnamefont {Babaei-Aghbolagh}}, \bibinfo {author} {\bibfnamefont {K.}~\bibnamefont {Babaei~Velni}}, \bibinfo {author} {\bibfnamefont {S.}~\bibnamefont {He}},\ and\ \bibinfo {author} {\bibfnamefont {Z.}~\bibnamefont {Pezhman}},\ }\href {https://doi.org/10.1007/JHEP07(2025)227} {\bibfield  {journal} {\bibinfo  {journal} {JHEP}\ }\textbf {\bibinfo {volume} {07}},\ \bibinfo {pages} {227}},\ \Eprint {https://arxiv.org/abs/2504.10361} {arXiv:2504.10361 [hep-th]} \BibitemShut {NoStop}%
\bibitem [{\citenamefont {Babaei-Aghbolagh}\ \emph {et~al.}(2026{\natexlab{b}})\citenamefont {Babaei-Aghbolagh}, \citenamefont {Babaei~Velni}, \citenamefont {He},\ and\ \citenamefont {Pezhman}}]{Babaei-Aghbolagh:2026vkm}%
  \BibitemOpen
  \bibfield  {author} {\bibinfo {author} {\bibfnamefont {H.}~\bibnamefont {Babaei-Aghbolagh}}, \bibinfo {author} {\bibfnamefont {K.}~\bibnamefont {Babaei~Velni}}, \bibinfo {author} {\bibfnamefont {S.}~\bibnamefont {He}},\ and\ \bibinfo {author} {\bibfnamefont {Z.}~\bibnamefont {Pezhman}},\ }\href {https://doi.org/10.1007/JHEP05(2026)037} {\bibfield  {journal} {\bibinfo  {journal} {JHEP}\ }\textbf {\bibinfo {volume} {05}},\ \bibinfo {pages} {037}},\ \Eprint {https://arxiv.org/abs/2602.03426} {arXiv:2602.03426 [hep-th]} \BibitemShut {NoStop}%
\end{thebibliography}%

\end{document}